\documentstyle[12pt,epsf]{article}
\newcommand{\be}{\begin{equation}}
\newcommand{\ee}{\end{equation}}
\newcommand{\bea}{\begin{eqnarray}}
\newcommand{\eea}{\end{eqnarray}}
\newcommand{\bean}{\begin{eqnarray*}}
\newcommand{\eean}{\end{eqnarray*}}

\begin{document}
\vspace{-1cm}
\noindent
\begin{flushright}
KANAZAWA-01-09\\
KUNS-1730
\end{flushright}
\vspace{10mm}
\begin{center}
{\Large \bf More about Kaluza-Klein Regularization
}
\vspace*{15mm}\\
\renewcommand{\thefootnote}{\alph{footnote}}
Tatsuo Kobayashi$^\dagger$
\footnote{E-mail: kobayash@gauge.scphys.kyoto-u.ac.jp}
and Haruhiko Terao$^{\dagger\dagger}$
\footnote{E-mail: terao@hep.s.kanazawa-u.ac.jp}
\vspace*{5mm}\\
$^\dagger$Department of Physics, Kyoto University\\ 
Kyoto 606-8502, Japan 
\vspace{2mm}\\
$^{\dagger\dagger}$Institute for Theoretical Physics, Kanazawa University\\
Kanazawa 920-1192, Japan
\end{center}
\vspace*{10mm}
\begin{abstract}
We study the so-called ``Kaluza-Klein regularization''.
We calculate  one-loop corrections to the Higgs mass
due to Kaluza-Klein modes explicitly in a model with 
SUSY breaking mass splitting between 
bosonic and fermionic modes.
We perform the proper time cutoff at $1/\Lambda^2$ and 
the KK level truncation at $\ell$.
It is shown that the finite result is obtained as long as
$\ell \gg R\Lambda$ for the compactification radius $R$.
\end{abstract}
\vspace*{20mm}
\noindent

\newpage
\pagestyle{plain}
\pagenumbering{arabic}
\setcounter{footnote}{0}

Recently, many efforts have been done to study 
supersymmetry (SUSY) breaking in extra dimensions 
as well as other phenomenologies in extra dimensions, 
and novel aspects have been found.
Among many interesting aspects, in Ref~.\cite{adpq}-\cite{dq} 
it has been found that one-loop radiative corrections 
to the Higgs mass are finite in the models with the Scherk-Schwarz 
SUSY breaking mechanism \cite{ss,antoniadis} and also in the models with
localized SUSY breaking 3-brane.
In such perturbative calculations, the so-called 
``Kaluza-Klein (KK) regularization'' is used, that is, 
infinite summation of all KK modes is taken first.

However, a doubt on the validity of the KK regularization 
has arisen \cite{gn,kim}.
It is claimed that the exchange between infinite sum and 
infinite integral would lead to an incorrect result.
In Ref~.\cite{gn}, the momentum cut-off $\Lambda$ is introduced and 
the KK-tower is truncated at the level $\ell$.
In addition, it is also claimed that
physically the level to contribute should be truncated around 
the momentum cut-off $\Lambda$.
In Ref~.\cite{gn}, such physical truncation is realized by a sharp cut,  
$\ell \approx \Lambda R$, 
where $R$ is the compactification radius.
Then, the quadratic divergence is found to appear 
in the Higgs mass corrections due to finite numbers of 
KK modes.

Moreover, there is a debate that the result of Ref.~\cite{gn} 
may be because of the sharp cut of the KK-tower, and 
such procedure spoils symmetries of the underlying theory.
Instead of such truncation, 
in Ref.~\cite{dgjq} the suppression by a Gaussian brane distribution is considered.
There the infinite numbers of KK modes are summed, but 
the couplings of higher modes are suppressed by a finite width of the brane.
Then the result is the same as the KK regularization, 
that is, the Higgs mass correction is finite.
However, it is pointed out also that there are other distributions 
leading to a linear sensitivity with the momentum cut-off $\Lambda$.
Also the Pauli-Villars regularization is discussed in 
Ref.~\cite{cp}.\footnote{
See also Ref.~\cite{msss,dkr}.}

In this letter, we consider the grounds for finiteness  of
the KK regularization.
We calculate explicitly the correction to the Higgs
mass due to the KK modes by performing both of the momentum cut-off
and the KK level truncation.
Our purpose is to show in which case the correction becomes insensitive 
to details of physics in the ultraviolet (UV) scale.
We discuss by using the proper time regularization for the UV divergence.
Because the proper time regularization does not spoil 
four-dimensional symmetries.
Also the proper time regularization is in a sense smooth compared with 
the sharp momentum cut-off.
Actually, the proper time regularization was used to 
derive the power-law behavior of gauge couplings 
due to KK modes \cite{power-law,kkmz,ktz}.
It was shown in Ref.~\cite{ktz} that transition to the power-law behavior 
at the compactification scale appears smoothly in the proper time regularization 
compared with other regularization schemes, 
although qualitatively same results for the power-law bahavior 
are obtained in different regularization schemes.
Hence, we use here the proper time regularization as a first trial. 
Then it will be shown that the finite result is obtained as long as
$\ell \gg R\Lambda$ for the compactification radius $R$.
This is seen also even when the sharp momentum cut-off 
is used \cite{gn}.

Suppose that we have the following bosonic and fermionic KK modes,
\begin{equation}
m_{b(n)} = {\sqrt \pi }{n+q \over R}, \qquad 
m_{f(n)} = {\sqrt \pi }{ n \over R},
\label{mass-sp}
\end{equation}
with $n = 0, \pm 1, \pm 2, \cdots$, where we have 
normalized the compactification radius $R$ with the factor $\sqrt \pi$ 
for convenience of the later calculations. 
For simplicity, we have assumed that only the bosonic modes have 
SUSY breaking masses.
However, the following discussions are applicable 
for more generic case.
On top of that, their coupling to the (zero-mode) Higgs field 
is denoted by $g$, which is assumed to be universal between 
bosonic and fermionic modes, and lower and higher KK modes.

In this setup, the fermionic contribution to 
the Higgs mass is proportional to the following integral,
\begin{equation}
g^2\sum^\infty_{n=-\infty} \int {d^4p \over (2\pi)^4} 
{1 \over p^2 + \pi n^2/R^2}.
\label{integral-1}
\end{equation}
Thus, how to calculate this and the corresponding bosonic 
integral is the point to be investigated.
Here we use the Scwinger representation, that is,  
we use the following identities:
\begin{equation}
\int^\infty_0 dt e^{-At} = {1 \over A}, \qquad 
\int {d^4p \over (2\pi)^4} e^{-tp^2} = {1 \over 16\pi^2 t^2}.
\end{equation}
Then the above integral (\ref{integral-1}) is written as
\begin{equation}
{g^2 \over 16\pi^2 } \sum^\infty_{n=-\infty} 
\int^\infty_0 dt ~{1 \over t^2} 
e^{-\pi tn^2/R^2} .
\end{equation}
Here we truncate the KK-modes at the level $\ell$ and
 put the cut-off $1/\Lambda^2$ for the proper time integral.
Then what should be evaluated is the following integral
\begin{equation}
I_f = {g^2 \over 16\pi^2 } \sum^\ell_{n=-\ell} 
\int^\infty_{1/\Lambda^2} dt ~{1 \over t^2} 
e^{-\pi tn^2/R^2} .
\end{equation}

Now, both of the summation and integral are finite, and 
we can exchange safely the summation and integration,
\begin{equation}
I_f = {g^2 \over 16\pi^2 }  
\int^\infty_{1/\Lambda^2} dt \sum^\ell_{n=-\ell} ~{1 \over t^2} 
e^{-\pi tn^2/R^2} .
\label{integral-2}
\end{equation}
Eq.~(\ref{integral-2}) includes the suppression factor 
$e^{-\pi tn^2/R^2}$ for higher KK modes $n \gg \Lambda R /\sqrt \pi$.
Thus, naively thinking, we can replace 
the finite summation by the infinite summation,
\begin{equation}
\int^\infty_{1/\Lambda^2} dt \sum^\ell_{n=-\ell} ~{1 \over t^2} 
e^{-\pi tn^2/R^2} \rightarrow 
\int^\infty_{1/\Lambda^2} dt \sum^\infty_{n=-\infty} ~{1 \over t^2} 
e^{-\pi tn^2/R^2}.
\label{replace-1}
\end{equation}
We shall discuss implication of this replacement later.
If the replacement (\ref{replace-1}) is allowed, 
the calculation is rather simple.
We use the Poisson summation formular, i.e. the 
modular transformation property of the $\theta_3(iA)$ function, 
\footnote{The replacement similar to Eq.~(\ref{replace-1}) 
and the Poisson summation formular are used in order to derive 
the power-law behavior of gauge couplings 
due to KK modes \cite{power-law,kkmz,ktz}.}
\begin{equation}
\theta_3(iA) \equiv \sum^\infty_{n=-\infty} e^{-\pi n^2 A} = {1 \over \sqrt A}
\sum^\infty_{m=-\infty} e^{-\pi A^{-1}m^2}= {1 \over \sqrt A}\theta_3(i/A),
\end{equation}
such that we obtain 
\begin{eqnarray}
I_f &=& {g^2 \over 16\pi^2 } \int^\infty_{1/\Lambda^2} dt ~{1 \over t^2} 
{R \over \sqrt t} \sum^\infty_{m=-\infty} e^{-\pi m^2 R^2/t},\\
&=& {g^2 \over 24\pi^2 } R\Lambda^3 + {g^2 \over 8\pi^{7/2} }
{C \over R^2} \zeta(3)
\label{integral-3}
\end{eqnarray}
where $C \equiv \int^\infty_{1/\Lambda^2} dy \sqrt y~e^{-y}$, 
and that is finite.
In particular, we have $C=\Gamma(3/2)$ in the limit 
$\Lambda^2 \rightarrow \infty$.
Thus we have the $\Lambda^3$ divergence for the fermionic 
modes only.

Similarly, we can calculate the contribution due to 
bosonic modes with the masses $(m_{b(n)})^2 = \pi (n+a)^2/R^2$.
The divergent term is exactly the same as Eq.~(\ref{integral-3}), 
although the finite part is different.
Thus, if the replacement (\ref{replace-1}) is allowed, 
we can deduce that the Higgs mass correction due to 
bosonic and fermionic KK modes with the 
mass spectrum (\ref{mass-sp}) is finite at the one-loop level 
of perturbative calculations.

In the above calculations, we have used the replacement
(\ref{replace-1}) of the finite number of summation 
by the infinite number of summation.
Such replacement by the infinite number of summation 
may be doubtable from the viewpoint of the sharp 
truncation.
Here we discuss more about the replacement (\ref{replace-1}).
First of all, each fermionic (bosonic) KK mode would 
have a quadratic divergent correction $\Lambda^2$ 
to the Higgs mass.
If we have $(2k +1)$ KK modes contributing to the 
Higss mass corrections, we have totally the corrections of 
the order of $(2k +1) \Lambda^2$.
We compare this naive estimation with the result
(\ref{integral-3}), 
whose divergence part is proportional to 
$\Lambda^3$.
That implies that in the proper time regularization and 
the replacement (\ref{replace-1}) the KK modes higher than 
$k \approx \Lambda R$ are decoupled effectively, 
but not by the explicitly sharp truncation, that is, 
the infinite summation seems not essential to obtain 
the finite result, but the summation over $\ell \gg k \approx \Lambda R$ is 
enough.
This result is also in agreement with the sharp cutoff case examined in
Ref.~\cite{gn}.

Next we examine the replacement (\ref{replace-1}).
For concreteness, we use the case with $a=1/2$.
For the moment, we require two points that 1) the positive and negative 
modes be treated on an equal footing and 2) the involved 
number of bosonic KK modes be the same 
as one of the corresponding fermionic modes in Eq.~(\ref{integral-2}).
Hence, the integral corresponding to $I_f$ (\ref{integral-2}) is obtained 
\begin{equation}
I_b = {g^2 \over 16\pi^2 }  
\int^\infty_{1/\Lambda^2} dt \sum^\ell_{n=-\ell} ~{1 \over 2t^2} 
\left[e^{-\pi t(n+1/2)^2/R^2} +e^{-\pi t(n-1/2)^2/R^2} \right].
\label{integral-b1}
\end{equation}
Now we estimate what we have added in the replacement (\ref{replace-1}).
We have added the following contribution for the fermionic modes:
\begin{equation}
I'_f = {g^2 \over 16\pi^2 }  
\int^\infty_{1/\Lambda^2} dt \sum^\infty_{n=\ell+1} ~{2 \over t^2} 
e^{-\pi tn^2/R^2} .
\label{integral-add-f}
\end{equation}
In a similar replacement of bosonic contribution, 
we have added the following contribution:
\begin{equation}
I'_b = {g^2 \over 16\pi^2 }  
\int^\infty_{1/\Lambda^2} dt \sum^\infty_{n=\ell +1} ~{1 \over t^2} 
\left[e^{-\pi t(n+1/2)^2/R^2} +e^{-\pi t(n-1/2)^2/R^2} \right].
\label{integral-add-b}
\end{equation}
Then the difference $I'_b-I'_f$ is written 
\begin{eqnarray}
I'_b - I'_f &=& {g^2 \over 16\pi^2 }  
\int^\infty_{1/\Lambda^2} dt \sum^\infty_{n=\ell+1} ~{1 \over t^2} 
\left[e^{-\pi t(n+1/2)^2/R^2} +e^{-\pi t(n-1/2)^2/R^2} - 
2e^{-\pi tn^2/R^2} \right] \nonumber \\
 &=& {g^2 \over 16\pi^2 }  
\int^\infty_{1/\Lambda^2} dt \sum^\infty_{n=\ell+1} ~{1 \over t^2} 
e^{-\pi tn^2/R^2}\left[ e^{-\pi t(n+1/4)/R^2}+ e^{\pi t(n-1/4)/R^2}
  -2
\right].
\label{level-1}
\end{eqnarray}
This differece $I'_b-I'_f$  contributes to the Higgs mass if 
$\ell$ is finite.
The terms in the bracket of Eq.~(\ref{level-1}) can be expanded
\begin{equation}
\left[ 1+1-2 -{\pi \over R^2} (n+{ 1\over 4}) t + 
{\pi \over R^2} (n-{ 1\over  4}) t + O(t^2)\right].
\label{level-2}
\end{equation}
The terms corresponding to the quadratic divergence $\Lambda^2$ 
seem to be cancelled, but the terms corresponding to 
the logarithmic divergence $\log \Lambda^2$  are 
not cancelled.
However, note that there is the suppression factor 
$e^{-\pi tn^2/R^2}$ in Eq.~(\ref{level-1}) with 
$n \geq \ell +1$ and $t \geq 1/\Lambda^2$.
Thus, if $\ell$ is large enough, such logarithmic divergence 
$\log \Lambda^2$ is decoupled.
That is consistent with the previous result.

On the other hand, around $\ell \approx R\Lambda$, such suppression 
factor does not work.
Then, we have the logarithmic divergence 
$\log \Lambda^2$ for each mode.
The divergence is enhanced 
by the number of relevant KK modes.
Thus, again, whether we have the divergence or not,  
depends on if we allow to take $\ell$ enough large.
We need not the infinite number for $\ell$.

The  splitting of the modes with 
the masses $n+1/2$ and $n-1/2$ in Eq.~(\ref{integral-b1}) might look 
artificial.
The reason for such representation is that we 
cared the edge modes with $n=\pm(\ell + 1/2)$.
However, now it is obvious that such higher modes 
have no contribution for sufficiently large $\ell$ because of 
the suppression factor $e^{-\pi \ell^2/(\Lambda R)^2}$.
Actually, 
the integral is written
\begin{equation}
I_b = {g^2 \over 16\pi^2 }  
\int^\infty_{1/\Lambda^2} dt \sum^\ell_{n=-\ell} ~{1 \over t^2} 
e^{-\pi t(n+1/2)^2/R^2},
\label{integral-b2}
\end{equation}
and the result is same, that is, we have the finite result 
for large $\ell \gg \Lambda R$.
Even in the case that we differ the levels for truncation between 
the fermionic and bosonic modes, say $\ell_f$ and $\ell_b$, 
we have the finite result for sufficiently large 
$\ell_f, \ell_b \gg R \Lambda$.

So far we have not been concerned about running of the gauge coupling 
(and Yukawa couplings, if any).
In extra dimensions the scale dependence is given by power 
law. 
We can incorporate this running coupling into the Higgs mass
corrections by examining the renormalization group (RG) equations for
$I_b$ (\ref{integral-b1}) and $I_f$ (\ref{integral-2}).
Indeed the finiteness is clearly seen in the RG point
of view.
Here we consider the RG equation in the Wilson sense,
which are given by
\begin{eqnarray}
\Lambda \frac{\partial I_f}{\partial \Lambda}
&=& {g^2(\Lambda) \over 16\pi^2 } \Lambda^2 \sum^\ell_{n = -\ell}
 e^{-\pi {n^2}/{(R\Lambda)^2}} 
\equiv {g^2(\Lambda) \over 16\pi^2 } \Lambda^2 {1 \over
  2}\varepsilon_f(R\Lambda),
\label{beta-1}\\
\Lambda \frac{\partial I_b}{\partial \Lambda}
&=& {g^2(\Lambda) \over 16\pi^2 } \Lambda^2 \sum^\ell_{n = -\ell} {1 \over 2}
\left[
e^{-\pi {(n+a)^2}/{(R\Lambda)^2}} + e^{-\pi {(n-a)^2}/{(R\Lambda)^2}}
\right]
\equiv {g^2(\Lambda) \over 16\pi^2 } \Lambda^2 {1 \over 2}\varepsilon_b(R\Lambda), 
\label{beta-2}
\end{eqnarray}
with $a=1/2$. Actually it is not necessary to give the beta function for the
gauge coupling in order to see the finiteness of the correction.
What concerns us is $\varepsilon_b(R\Lambda)-\varepsilon_f(R\Lambda)$.
If this difference vanishes for the region $R\Lambda > 1$, 
the corrections to the Higgs mass would be UV insensitive and finite.

For example we take $\ell = 10$.
Fig.~1 shows $\varepsilon_b(x)$ and $\varepsilon_f(x)$.
Behaviors of the two curves are almost same for $x > 0.5$.
In the region with $x<10$, i.e. $R\Lambda < \ell$,  
the curve of $\varepsilon_{b,f}(x)$ behaves linearly.
This behavior is consistent with the previous result 
(\ref{integral-3}) that 
the divergence behaves like $\Lambda^3$ for $R\Lambda \ll \ell$.
For large $x$, $\varepsilon_{b,f}(x)$ goes close to 
the constant $2(2\ell +1)$.

\begin{figure}[htb]
\begin{center}
\epsfxsize=0.5\textwidth
\leavevmode
\epsffile{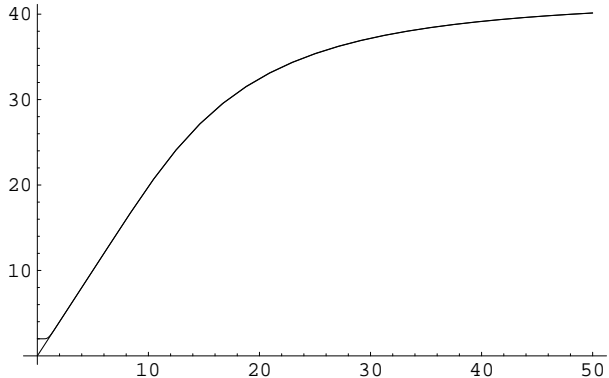}
\caption{Power law behavior of $\varepsilon_b(x)$ and $\varepsilon_f(x)$.}
\end{center}
\end{figure}

Fig.~2 shows the difference 
$\varepsilon_f(R\Lambda)-\varepsilon_b(R\Lambda)$.
The difference damps rapidly above $x \sim 1$.
That implies that for $1 \leq R\Lambda < \ell$, 
no correction arises to the Higgs mass.
Therefore, as long as we keep the KK level truncation as
$R\Lambda \ll \ell$, the correction is finite and 
of $O(1/R^2)$.
It was mentioned above that we need large $\ell \gg \Lambda R$ 
in order to obtain the finite result.
This numerical calculation shows how large $\ell$ is necessary, and 
$\Lambda R/\ell < 0.9$ seems sufficient, although the explicit value 
0.9 has no serious meaning.
That suggests a huge gap between $\ell$ and $\Lambda R$ 
is not required.
There is a small lump starting from $x \approx 10$, 
i.e. $\Lambda R \approx \ell$.
That corresponds to the sum of logarithmic divergences, 
which is mentioned in the level-by-level calculation 
around Eq.~(\ref{level-2}), and that is also 
consistent with the calculation in Ref.~\cite{gn} 
showing the presence of divergence,
where the level truncation is taken around  $\ell \approx \Lambda R$.
The difference $\varepsilon_f(R\Lambda)-\varepsilon_b(R\Lambda)$
behaves as $x^{-2}$ at $x \gg l$. This implies that the corrections
depend on the UV cut-off $\Lambda$ logarithmically (with constant $g$),
when we truncate the KK modes at much lower scale than $\Lambda$.

\begin{figure}[htb]
\begin{center}
\epsfxsize=0.5\textwidth
\leavevmode
\epsffile{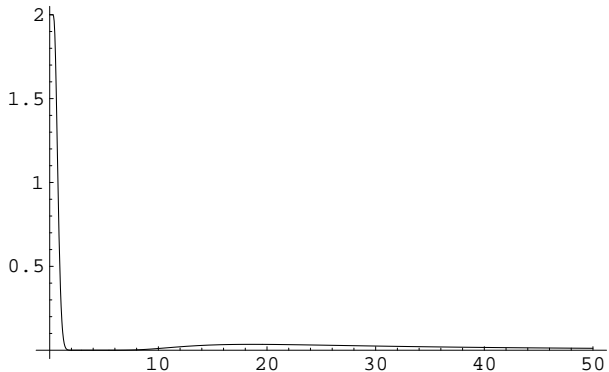}
\caption{$\varepsilon_f(x)-\varepsilon_b(x)$.}
\end{center}
\end{figure}

We have calculated numerically the case with $a=1/2$.
We can easily extend to other cases, e.g. for other values of $a$, 
\begin{eqnarray}
I_b &=& {g^2 \over 16\pi^2 }  
\int^\infty_{1/\Lambda^2} dt \sum^\ell_{n=-\ell} ~{1 \over t^2} 
e^{-\pi t(n+a)^2/R^2}, 
\label{integral-b3}\\
\Lambda \frac{\partial I_b}{\partial \Lambda}
&=& {g^2 \over 16\pi^2 } \Lambda^2 \sum^\ell_{n = -\ell} 
e^{-\pi {(n+a)^2}/{(R\Lambda)^2}} 
\equiv {g^2 \over 16\pi^2 } \Lambda^2 {1 \over 2}\varepsilon_b(R\Lambda). 
\label{beta-3}
\end{eqnarray}
For any value of $a$, the difference 
$\varepsilon_f(R\Lambda) - \varepsilon_b(R\Lambda)$ behaves like
Fig.~2 in the region with $x< \ell$.
That implies for $\ell > \Lambda R$ the correction to the Higgs mass 
is finite and of $O(1/R^2)$.
Furthermore, 
in $I_b$ of Eq.~(\ref{integral-b3}) 
we can replace the truncated level $\ell$ by $\ell_b$, which is independent of 
the truncated level of fermionic modes $\ell$.
Even in this case, we have the same behavior of 
$\varepsilon_f(R\Lambda) - \varepsilon_b(R\Lambda)$ 
in the region with $x< \ell, \ell_b$.

We have used the proper time regularization.
However we can apply also other regularization schemes,
e.g. the sharp momentum cut-off.
Explicitly we examine the beta function
\begin{equation}
\Lambda {\partial I_f \over \partial \Lambda }
= {g^2 \over 16\pi^2 } \sum_{n=-\ell}^\ell 
{\Lambda^4 R^2 \over (\Lambda R)^2 + \pi
  n^2}
\equiv {g^2 \over 16\pi^2 } \Lambda^2 {1 \over 2} 
\varepsilon_f(\Lambda R)
\end{equation}
and the similar function for the bosonic contributions.
Then the power law behavior does not depend on the regularization
scheme, except for the KK threshold corrections \cite{ktz}.
Therefore we obtain qualitatively the same results 
for the finiteness of the corrections.

Also similar discussions seem useful not only for the model with 
the mass spectrum (\ref{mass-sp}), but more generic case.
For any other mass spectra, we can define 
the difference $\varepsilon_f(R\Lambda)-\varepsilon_b(R\Lambda)$
like Eqs.~(\ref{beta-1}) and (\ref{beta-2}).
Then it can be used as an index for the presence of 
quadratic divergences.

To summarize, we have calculated explicitly the Higgs mass correction 
due to KK modes.
The momentum cut-off $\Lambda$ and the level of truncation $\ell$ 
have been included.
We have used the proper time regularization, 
which decouples effectively higher KK modes $n > \Lambda R$, 
and that would fit the philosophy of Ref.~\cite{gn} that such 
higher modes should be decoupled.
However, the decoupling by the proper time regularization is 
smooth compared with the sharp cut of Ref.~\cite{gn}.
Our results are in order.
Finiteness or the appearance of 
divergences does not depend on whether we put a finite truncation of 
the KK modes or we sum the infinite number of KK modes, 
but it depends on where we take the finite truncation $\ell$.
If we are allowed sufficiently large truncation 
compared with the momentum cut-off $\ell \gg \Lambda R$, 
we obtain the finite result.
Numerical study shows we need a large truncation $\ell$, but not 
a huge gap between $\ell$ and $\Lambda R$.
On the other hand, if we put $\ell \approx \Lambda R$, 
the divergence appears.
That might be rather obvious.
Because truncating around the momentum cut-off means 
we would see how to truncate the modes.
Therefore, finiteness depends if theory allows $\ell \gg \Lambda R$.
For the spectrum (\ref{mass-sp}), it might be artificial to assign 
the bosonic state of the mass $m_{b(n)} = {\sqrt \pi }{(n+q)/R}$ 
with the fermionic state of the mass 
$m_{f(n)} = {\sqrt \pi }{ n/R}$.
The corresponding bosonic state might be the state with 
$m_{b(n)} = {\sqrt \pi }{(n-1+ q)/R}$.
For example, suppose 
a theory, which should be irrelevant to such artificial truncation of 
the edge.
In this case one has to take $\ell \gg \Lambda R$, such that 
low energy physics becomes insensitive to how to treat 
such edge.
The concept of locality in extra dimensions, which are 
discussed in Refs.~\cite{adpq}-\cite{dq}, might forbid 
the truncation of KK modes around the cut-off.
On the other hand, suppose a theory, where 
the cut-off $\Lambda$ has a real meaning, e.g. the string scale.
In this case, we have to truncate at $\ell \approx \Lambda R$, 
above which new modes might appear.
For such case, we have to know initial conditions 
at $\Lambda$ to discuss low energy physics.

The radiative corrections in the string theory with a large radius of 
compactification may be well approximated by using the effective field 
theroy with infinite tower of the KK modes \cite{gr}. 
There the corrections are represented  by a kind of proper time cutoff,
with which the KK modes heavier than the string scale are decoupled. 
Therefore the correction to the higgs mass is supposed to be insensitive 
to the string scale. 

Finite results for the Higgs mass correction due to KK modes 
have been found first for models with SUSY breaking by 
the Scherk-Schwarz mechanism Refs.~\cite{adpq} -\cite{dq}.
However, what we have studied is the calculations of the 
correction under the mass spectrum (\ref{mass-sp}).
Whether the spectrum is obtained by 
the Scherk-Schwarz mechanism  or other SUSY breaking mechanisms, 
is irrelevant to the calculations.
Thus, finite results are generic for models with 
the mass spectrum.
To repeat, the key-point is whether models allow 
to take large truncation $\ell \gg \Lambda R$.

Our discussions for finiteness are valid at the one-loop level of 
perturbation theory.
The two-loop level \cite{two-loop} and more are beyond the scope.

\section*{Acknowledgments}

The authors would like to thank D.M.~Ghilencea, Y.~Murakami, H.~Nakano 
and K.~Yoshioka for useful discussions.

\end{document}